\documentclass[twocolumn,amsmath,amssymb]{revtex4}

\usepackage{graphicx}
\usepackage{dcolumn}
\usepackage{bm}
\begin{document}


\title{Isospin Dynamics in Heavy Ion Collisions}
\author{M.Di Toro}
\affiliation{Laboratori Nazionali del Sud INFN, Via S.Sofia 62, 95123 
Catania, Italy,\\
and Physics-Astronomy Dept., University of Catania,\\
E-mail: ditoro@lns.infn.it}

\begin{abstract}
Some isospin dynamics results in heavy ion collisions from low 
to relativistic energies obtained through transport approaches, largely
inspired by David M.Brink, are rewieved.
At very low energies, just above the Coulomb barrier, the stimulating 
implications of the prompt dipole radiation in dissipative collisions
of ions with large isospin asymmetries are discussed.   
We pass then to the very rich phenomenology of isospin effects on heavy 
ion reactions
at intermediate energies (few $A~GeV$ range). We show that it can allow
a ``direct'' study of the covariant structure of the isovector interaction
in the hadron medium. We work within a relativistic transport frame,
beyond a cascade picture, 
 consistently derived from effective Lagrangians, where isospin effects
are accounted for in the mean field and collision terms.
Rather sensitive observables are proposed from collective flows
(``differential'' flows) and from pion/kaon production ($\pi^-/\pi^+$, 
$K^0/K^+$ yields). For the latter point relevant non-equilibrium effects
are stressed.
The possibility of the transition to a mixed hadron-quark phase, 
at high baryon and isospin density, is finally suggested. Some signatures
could come from an expected ``neutron trapping'' effect.
The importance of {\it violent} collision experiments with radioactive
beams, from few $AMeV$ to few $AGeV$, is stressed.
\end{abstract}
\maketitle

\section{Introduction: the David Legacy}

It is a great pleasure and honor to contribute to this special day devoted to
celebrate the 75th anniversary of David Brink. After many years of 
acquaintance and collaboration I would like to state a few points I mostly got
from him: 1.) The Mean Field ``moves'' the world, particularly true in
nuclear dynamics, as I will show in examples discussed here; 2.) The role
of relativity in nuclear structure and reactions; 3.) Play always a deep 
attention to the suggestions of young people; 4.) Follow a ``sensible''
behaviour towards the political choices, i.e. try to change something only
when you can count on something better.
Here of course I will focus on the physics part showing a series of results
broadly inspired by the David ideas. 

In the last years the isospin dynamics has gained a lot of interest, as we can
see from the development of new heavy ion facilities (radioactive beams),
for the unique possibilities of probing the isovector in medium interaction
far from saturation, relevant for the structure of unstable elements as well 
as for nuclear astrophysics see the recent reviews \cite{bao,baranPR}.

Here I will show some selected results of the mean field transport approach
in a wide energy range, from few $AMeV$ to few $AGeV$, in non-relativistic 
and relativistic frames. At low energies I will discuss the isospin 
equilibration in dissipative collisions, fusion and deep-inelastic, through
a related observable, the Prompt Collective Dipole Radiation. At high
energies I will shortly present isospin effects on collective flows, on 
particle production
and finally on the transition to a mixed hadron-quark phase at high baryon
density.

\section{The Prompt Dipole $\gamma$-Ray Emission}

The possibility of an entrance channel bremsstrahlung dipole radiation
due to an initial different N/Z distribution was suggested at the beginning
of the nineties \cite{ChomazNPA563,BortignonNPA583}, largely inspired by
David discussions. At that time a large debate was present on the disappearing
of $Hot~Giant~Dipole~Resonances$ in fusion reactions. David was suggesting
the simple argument that a $GDR$ needs time to be built in a hot compound
 nucleus, meanwhile the system will cool down by neutron emission and the
$GDR$ photons will show up at lower temperature. The natural consequence 
suggested in \cite{ChomazNPA563} was that we would expect a new
dipole emission, in addition to the statistical one, if some pre-compound 
collective dipole mode is present.
After several experimental evidences, in fusion as well as in deep-inelastic
reactions \cite{FlibPRL77,CinNC111,PierrouEPJA16,AmoPRC29,PierrouPRC71} we have
now a good understanding of the process and stimulating new perspectives
from the use of radioactive beams.

During the charge equilibration process taking place
 in the first stages of dissipative reactions between colliding ions with
 different N/Z
ratios, a large amplitude dipole collective motion develops in the composite 
dinuclear system, the so-called dynamical dipole mode. This collective dipole 
gives rise to a prompt $\gamma $-ray emission which depends:  
 i) on the absolute 
value of the intial dipole moment
\begin{eqnarray}
&&D(t= 0)= \frac{NZ}{A} \left|{R_{Z}}(t=0)- {R_{N}}(t=0)\right| =  \nonumber \\
&&\frac{R_{p}+R_{t}}{A}Z_{p}Z_{t}\left| (\frac{N}{Z})_{t}-(\frac{N}{Z}){p}
\right|, 
\label{indip}
\end{eqnarray}
being ${R_{Z}}$ and ${R_{N}}$ the
center of mass of protons and of neutrons respectively, while R$_{p}$ and
R$_{t}$ are the
projectile and target radii; ii) on the fusion/deep-inelastic dynamics;
 iii) on the symmetry term, below saturation, that is acting as a restoring 
force.

A detailed description is obtained in a
microscopic approach based on semiclassical transport equations,
of Vlasov type, introduced in the nuclear dynamics in collaboration
with David \cite{BrinkNPA372},
where mean field and two-body collisions are treated in a
self-consistent way, see details in \cite{BaranNPA600}. Realistic
effective interactions of Skyrme type are used. The numerical
accuracy of the transport code has been largely improved in order
to have reliable results also at low energies, just above the
threshold for fusion reactions 
\cite{CabNPA637,BaranNPA679}. The
resulting physical picture is in good agreement with quantum
Time-Dependent-Hartree-Fock calculation \cite{SimenPRL86}. In
particular we can study in detail how a collective dipole
oscillation develops in the entrance channel \cite{BaranNPA679}.

First, during the ${\it approaching~phase}$, the two partners,
overcoming the Coulomb barrier, still keep their own response.
Then follows a {\it dinuclear phase} where the
relative motion energy, due to the nucleon exchange, is converted 
in thermal motion and in 
the collective
energy of the dinuclear mean field.
In fact the composite system is not
fully  equilibrated and  manifests, as a whole, a large
amplitude  dipole collective motion. Finally thermally
equilibrated reaction products are formed, with consequent
statistical particle/radiation emissions.

\begin{figure}
\includegraphics[scale=0.37]{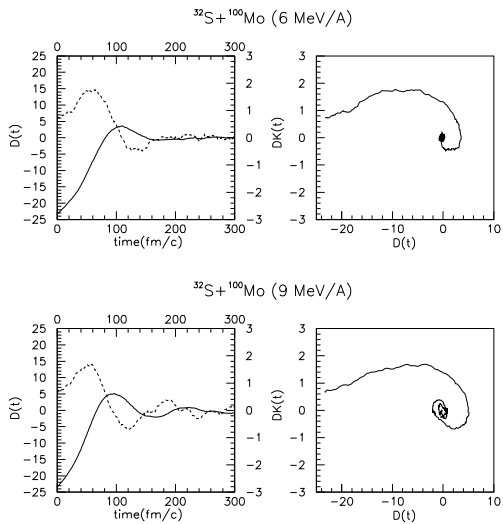}
\caption{
The time evolution of the dipole mode
in $r-$space $D(t)$ (solid lines) and $p-$space $DK(t)
$
(dashed lines, in
$fm^{-1}$) and the correlation $DK(t) -  D(t)$
at incident energy of $6AMeV$ and $9AMeV$ for $b= 2 fm$.}
\label{spiral}
\end{figure}

\begin{figure}
\includegraphics[scale=0.40]{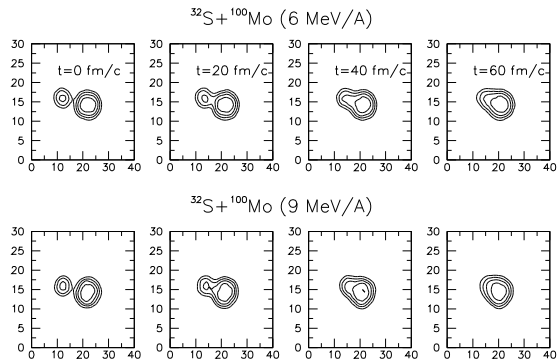}
\caption{Density plots of the neck dynamics for the $^{32}$S +$^{100}$Mo 
system at incident energy of $6AMeV$ and $9AMeV$.}
\label{fusdyn}
\end{figure}

\begin{table*}[hbt]
\caption{\it The percent increase of the intensity in the linearized 
$\gamma $-ray
spectra at the compound nucleus GDR energy region (the energy integration
 limits are given in the parenthesis), the compound
nucleus excitation energy, the initial dipole moment
$D(t=0)$ and the initial mass asymmetry $\Delta$ for each reaction.}
\begin{ruledtabular}
\begin{tabular}{lcccccccccc}
Reaction  &  Increase (\%) & E$^{*}$(MeV)& $D(t=0)$ (fm) & $\Delta$ & 
 Ref \\
\hline
$^{40}$Ca+$^{100}$Mo & 16 (8,18)& 71 &22.1 & 0.15 & \cite{FlibPRL77} \\
$^{36}$S+$^{104}$Pd & & 71 & 0.5 & 0.17   \\
\hline
$^{16}$O+$^{98}$Mo &  36 (8,20)& 110 &8.4 & 0.29 & \cite{CinNC111} \\
$^{48}$Ti +$^{64}$Ni & & 110 & 5.2 & 0.05 \\
\hline
$^{32}$S+$^{100}$Mo &  1.6 $\pm 2.0$ (8,21)& 117 &18.2 & 0.19 & 
\cite{PierrouPRC71}\\
$^{36}$S+$^{96}$Mo &  & 117 &1.7 & 0.16 \\
\hline
$^{32}$S+$^{100}$Mo &  25 (8,21)& 173.5 &18.2 & 0.19 & \cite{PierrouEPJA16} \\
$^{36}$S+$^{96}$Mo &  & 173.5 &1.7 & 0.16  
\end{tabular}
\end{ruledtabular}
\end{table*}

We present here some results for the $^{32}$S +$^{100}$Mo ($N/Z$ asymmetric)
reaction at $6~and~9~A MeV$, recently studied vs. the 
``symmetric'' $^{36}$S +$^{96}$Mo
counterpart in ref.\cite{PierrouPRC71}.
In Fig.\ref{spiral} (left columns) we draw the time
evolution for $b= 2 fm$ of the dipole moment
in the $r$-space (solid lines), $D(t)= \frac{NZ}{A} X(t)$ and in 
$p-$space (dashed lines), $DK(t)= \Pi/\hbar$, where 
$\Pi= \frac{NZ}{A}(\frac{P_{p}}{Z}-\frac{P_{n}}{N})$, with $P_{p}$
($P_{n}$) center of mass in momentum space for protons (neutrons),
is just the canonically conjugate momentum of the $X$ coordinate,
see \cite{BaranNPA679,SimenPRL86,BaranPRL87}. On the right hand side 
columns we show the 
corresponding correlation $DK(t) -  D(t)$
in the phase space.
We choose the origin of time at the
beginning of the {\it dinuclear} phase. The nice "spiral-correlation"
clearly denotes the collective nature
 of the mode.
From Fig.\ref{spiral} we note that the "spiral-correlation" starts when
the initial dipole moment $D(t= 0)$, the geometrical value at the
touching point, is already largely quenched. This is the reason
why the dinucleus dipole yield is not simply given by the "static"
estimation but the reaction dynamics has a large influence on it.

A clear energy
dependence of the dynamical dipole mode is evidenced
with a net increase when we pass
from $6AMeV$ to $9AMeV$. A possible explanation
of this effect is due to the fact that at lower energy, just above the Coulomb
barrier, a longer
transition to a dinuclear configuration is required 
which hinders the isovector
collective response. From Fig.\ref{fusdyn}
a  slower dynamics of the neck
during the first $40fm/c - 60fm/c$ from the touching
configuration is observed at $6~AMeV$. When the collective dipole response 
sets in the charge is already partially equilibrated via random nucleon 
exchange.

\begin{figure}
\includegraphics[scale=0.37]{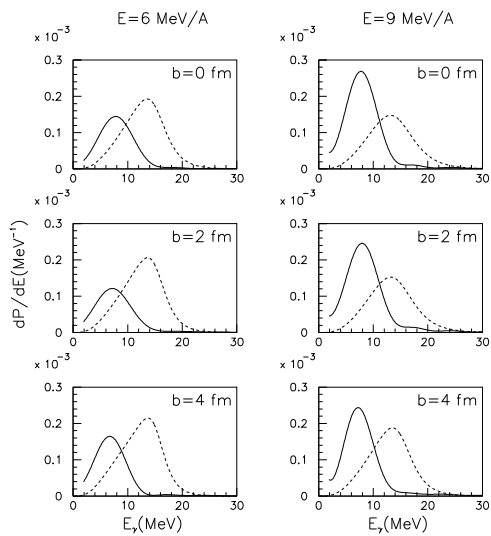}
\caption{The bremsstrahlung spectra for the $^{32}$S +$^{100}$Mo system
at incident energy of $6AMeV$ and $9AMeV$ (solid line) and the first step
statistical spectrum (dashed line) for three
impact parameters.}
\label{spectra}
\end{figure}

The bremsstrahlung spectra shown in Fig.\ref{spectra} support this 
interpretation.
 
In fact from the dipole evolution given from the Vlasov transport
we can directly
apply a bremsstrahlung {\it ("bremss")}  approach
\cite{BaranPRL87} to estimate the ``direct'' photon emission probability
($E_{\gamma}= \hbar \omega$):
\begin{equation}
\frac{dP}{dE_{\gamma}}= \frac{2 e^2}{3\pi \hbar c^3 E_{\gamma}}
 |D''(\omega)|^{2}  \label{brems},
\end{equation}
where $D''(\omega)$ is the Fourier transform of the dipole acceleration
$D''(t)$. We remark that in this way it is possible
to evaluate, in {\it absolute} values, the corresponding pre-equilibrium
photon emission.
In the same Fig.\ref{spectra} we show statistical $GDR$ emissions from the
final excited residue. We see that at the higher energy the prompt emission
represents a large fraction of the total dipole radiation. 

In the Table we report the present status of the Dynamical Dipole data, 
obtained from fusion reactions..
We note the dependence of the extra strength on the interplay between
initial dipole moment and initial mass asymmetry: this clearly indicates
the relevance of the fusion dynamics.

We must add a 
couple of comments of interest for the experimental selection of the Dynamical
Dipole: i) The centroid is always shifted to lower energies (large 
deformation of the dinucleus); ii) A clear angular anisotropy should be present
since the prompt mode has a definite axis of oscillation 
(on the reaction plane) at variance with the statistical $GDR$.
At higher beam energies we expect a further decrease of the direct dipole 
radiation for two main reasons both due to the increasing importance of hard
NN collisions: i) a larger fast nucleon emission that will equilibrate the 
isospin before the collective dipole starts up; ii) a larger damping of the 
collective mode. This has been observed in ref.\cite{AmoPRC29} and more exps. 
are planned \cite{PierrouLNS}.

Before closing I would like to note two interesting developments for future
experiments with radioactive beams: 
\begin{itemize}
\item{The prompt dipole radiation represents a nice cooling mechanism on the
fusion path. It could be a way to pass from a {\it warm} to a {\it cold}
fusion in the synthesis of heavy elements with a noticeable increase of the
{\it survival} probability, \cite{luca04}.}
\item{The use of unstable neutron rich projectiles would largely increase the
effect allowing a detailed study of the symmetry potential, below saturation,
responsible of the restoring force of the dipole oscillation \cite{spiral2}}
\end{itemize}

\section{Isospin Physics in a Covariant Approach}
We move now to a relativistic framework for the description of
the isovector part of the effective interaction.
I will focus then the  attention on relativistic heavy ion collisions, that
provide a unique terrestrial opportunity to probe the in-medium nuclear
interaction in high density and high momentum regions. 
An effective Lagrangian approach to the hadron interacting system is
extended to the isospin degree of freedom: within the same frame equilibrium
properties ($EoS$, \cite{qhd}) and transport dynamics 
(\cite{KoPRL59,GiessenRPP56}) can be consistently derived.
Within a covariant picture of the nuclear mean field, 
 for the description of the symmetry energy at saturation
($a_{4}$ parameter of the Weizs\"{a}ecker mass formula)
  (in a sense equivalent to the $a_{1}$ parameter for the 
 iso-scalar part), extracted in the range from 28 to 36 MeV, 
there are different 
possibilities: (a) considering only the Lorentz vector $\rho$ mesonic field, 
and (b) both, the vector $\rho$ (repulsive) and  scalar 
$\delta$ (attractive) effective 
fields \cite{kubis,liu,gait04}. The latter corresponds to the two 
strong effective $\omega$ (repulsive) and 
$\sigma$ (attractive) mesonic 
fields of the iso-scalar sector.
We get a transparent form \cite{liu,baranPR}: 
\begin{equation}
E_{sym} = \frac{1}{6} \frac{k_{F}^{2}}{E_{F}^*} + 
\frac{1}{2}
\left[ f_{\rho} - f_{\delta}\left( \frac{m^{*}}{E_{F}^*} \right)^{2}
\right] \rho_{B}, 
\label{esym3}
\end{equation}
with $E_{F}^* \equiv \sqrt{k_{F}^{2} + {m^*}^2}$.

Once the $a_{4}$ empirical value is fixed from the  
$\rho-\delta$ balance, 
important effects at supra-normal densities appear due to the 
introduction of the effective $\delta$ field. In fact 
the presence of an isovector scalar field is increasing the repulsive 
$\rho$-meson contribution at high baryon densities \cite{liu,baranPR} 
via a pure relativistic mechanism, due to the different Lorentz 
properties of these fields 
(the vector $\rho$ field grows with baryon density whereas the scalar 
$\delta$ field is suppressed by the scalar density).
Dynamical, non-equilibrium, effects can be more sensitive to such ``fine
structure'' of the isovector interaction. We will see the vector couplings give
$\gamma$-boosted Lorentz forces,  and so we expect
a larger isospin dependence of the high energy nucleon propagation. 
Moreover, the scalar $\delta$ field naturally 
leads to an effective ($Dirac$) mass splitting between protons and 
neutrons \cite{baranPR,kubis,liu,gait04,dbhf,diracmass}, with influence on 
nucleon emissions and flows \cite{msuria,RizzoPRC72}. 
In order to explore the symmetry energy at supra-normal densities one  
has to select signals directly emitted from the early non-equilibrium 
high density 
stage of the heavy ion collision.
A transverse momentum analysis is important in order
to select the high density source \cite{dani,gait01}. 
The description of the mean field is important, since nucleons and resonances
are {\it dressed} by the self-energies. This  will
directly affect the energy balance (threshold and phase space) 
of the inelastic channels. 

\subsection*{Relativistic Transport}
The starting point is
a simple phenomenological version of the Non-Linear (with respect to the 
iso-scalar, Lorentz scalar $\sigma$ field) Walecka model 
which corresponds 
to the 
Hartree or Relativistic Mean Field ($RMF$) approximation within the 
Quantum-Hadro-Dynamics \cite{qhd}. 
According to this model the baryons (protons and neutrons) are described by an 
effective Dirac equation $(\gamma_{\mu}k^{*\mu}-M^{*})\Psi(x)=0$, whereas the 
mesons, which generate the classical mean field, are characterized by 
corresponding covariant equations of motion.
The presence of the hadronic medium modifies the masses and momenta 
of the hadrons, i.e. $M^{*}=M+\Sigma_{s}$ (effective masses), 
 $k^{*\mu}=k^{\mu}-\Sigma^{\mu}$ (kinetic momenta), where we have introduced
the scalar and vector self-energies $\Sigma_{s},~\Sigma^{\mu}$ . 
For asymmetric matter the self-energies are different for protons and 
neutrons, depending on the isovector meson contributions. 
We will call the 
corresponding models as $NL\rho$ and $NL\rho\delta$, respectively, and
just $NL$ the case without isovector interactions. We will show also
some results with Density Dependent couplings, in order to probe effects which
go beyond the $RMF$ picture.
For the more general $NL\rho\delta$ case  
the self-energies 
of protons and neutrons read:
\begin{eqnarray}
\Sigma_{s}(p,n)& = & - f_{\sigma}\sigma(\rho_{s}) \pm f_{\delta}\rho_{s3}, 
\nonumber \\
\Sigma^{\mu}(p,n)& = &f_{\omega}j^{\mu} \mp f_{\rho}j^{\mu}_{3},
\label{selfen}
\end{eqnarray}
(upper signs for neutrons),
 where $\rho_{s}=\rho_{sp}+\rho_{sn},~
j^{\alpha}=j^{\alpha}_{p}+j^{\alpha}_{n},\rho_{s3}=\rho_{sp}-\rho_{sn},
~j^{\alpha}_{3}=j^{\alpha}_{p}-j^{\alpha}_{n}$ are the total and 
isospin scalar 
densities and currents and $f_{\sigma,\omega,\rho,\delta}$  are the coupling 
constants of the various 
mesonic fields. 
$\sigma(\rho_{s})$ is the solution of the non linear 
equation for the $\sigma$ field \cite{liu,baranPR}.

For the description of heavy ion collisions we solve
the covariant transport equation of the Boltzmann type 
 \cite{KoPRL59,GiessenRPP56}  within the 
Relativistic Landau
Vlasov ($RLV$) method, phase-space Gaussian test particles \cite{FuchsNPA589},
and applying
a Monte-Carlo procedure for the hard hadron collisions.
The collision term includes elastic and inelastic processes involving
the production/absorption of the $\Delta(1232 MeV)$ and $N^{*}(1440
MeV)$ resonances as well as their decays into pion channels,
 \cite{huber,FeriniNPA762}.

A relativistic kinetic equation can be obtained from nucleon Wigner Function
dynamics derived from the effective Dirac equation \cite{GiessenRPP56}.
The neutron/proton Wigner functions are expanded in terms of
components with definite transformation properties. Consistently with 
the effective fields
included in our minimal model one can limit the expansion to 
scalar and vector parts 
$
{\hat F}^{(i)}(x,p) = F_S^{(i)}(x,p) + \gamma_\mu F^{{(i)}\mu}(x,p),~~~i=n,p.
$
We get after some algebra
a transport equation of Vlasov type 
for the scalar part $f_i(x,p^{* \mu})
 \equiv F_S^i/M_i^*$; 
$\{p_{\mu i}^{*}\partial^\mu + [p_{\nu i}^{*} F_i^{\mu\nu} + 
 M_i^*(\partial^\mu M_i^*)] {\partial_\mu}^{p^*} \} f_i(x,p^{{*}\mu})~=~0,
$
with the field tensors
$
F_i^{\mu\nu} \equiv \partial^\mu p_i^{{*}\nu} + \partial^\nu p_i^{{*}\mu}. 
$
The trajectories of test particles obey to the following equation
of motion:
\begin{eqnarray}
&&\frac{d}{d\tau}x_i^\mu=\frac{p^*_i(\tau)}{M^*_i(x)}~,
\nonumber \\
&&\frac{d}{d\tau}p^{*\mu}_i=\frac{p^*_{i\nu}(\tau)}{M^*_i(x)}
{F}_i^{\mu\nu}\left(x_i(\tau)\right)+\partial^\mu M^*_i(x)~.
\label{eqmot}
\end{eqnarray}

In order to have an idea of the dynamical effects of the
covariant nature of the interacting fields, we
write down, with some approximations, the ``force'' acting on a particle. 
Since we are interested in isospin contributions we will take into account 
only the isovector part of the interaction \cite{GrecoPLB562}:
\begin{eqnarray}\label{force}
\frac{{d\vec p}^{\,*}_i}{d\tau}& = &\pm f_{\rho} \frac{p_{i\nu}}{M^*_i}
\left[\vec\nabla J_{3}^{\nu}-\partial^\nu\vec{J_3} \right]
\mp f_\delta \nabla \rho_{S3} \nonumber \\
& \approx & 
\pm f_{\rho} \frac{E^*_i}{M^*_i}
\vec\nabla \rho_{3}
\mp f_\delta \vec\nabla \rho_{S3}, ~~(p/n) 
\end{eqnarray}
The Lorentz force (first term of Eq.(\ref{force}) shows a 
$\gamma=\frac{E^*_i}{M^*_i}$ boosting of the vector coupling, while 
from the second term we expect a $\gamma$-quenched $\delta$ contribution.
We remark that the Lorentz-like force is absent in the non-relativistic
Vlasov transport equation discussed before. This nicely shows the qualitative
different dynamics of a fully relativistic approach. You cannot get it just
inserting a relativistic kinematics in the classical transport equations.

\section{Collective Flows}
The flow observables can be seen respectively as the
first and second coefficients of a Fourier expansion of the
azimuthal distribution \cite{OlliPRD46}:
$\frac{dN}{d\phi}(y,p_t) \approx 1+2V_1cos(\phi)+2V_2cos(2\phi)$
where $p_t=\sqrt{p_x^2+p_y^2}$ is the transverse momentum and $y$
the rapidity along beam direction. 
The transverse flow can be also
expressed as: $V_1(y,p_t)=\langle \frac{p_x}{p_t} \rangle$.
The sideward (transverse) flow is a deflection of forwards and backwards 
moving particles, within the reaction plane. 
The second  coefficient of the expansion defines the elliptic
flow given by
$V_2(y,p_t)=\langle \frac{p_x^2-p_y^2}{p_t^2} \rangle$.
\begin{figure}
\begin{center}
\includegraphics[angle=-90,scale=0.35]{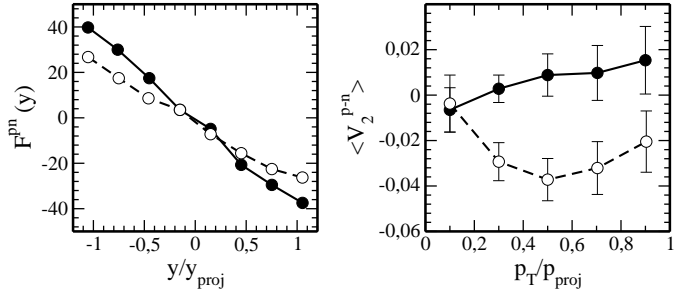}
\caption{Differential neutron-proton flows for the $^{132}Sn+^{124}Sn$
reaction at 1.5 AGeV (b=6fm) from the three different models for the
isovector mean fields.
Top: Transverse Flows. Bottom: Elliptic Flows.
Full circles and solid line: $NL\rho\delta$.
Open circles and dashed line: $NL\rho$.
} 
\label{flows}
\end{center}
\end{figure}
It measures the competition between in-plane and out-of-plane emissions. 
 The sign of $V_2$ indicates the azimuthal anisotropy of emission:
particles can be preferentially emitted either in the reaction
plane ($V_2>0$) or out-of-plane ($squeeze-out,~V_2<0$)
\cite{OlliPRD46,DanielNPA673}. The $p_t$-dependence of $V_2$ is
very sensitive to the high density behavior of the $EoS$ since highly
energetic
particles ($p_t \ge 0.5$) originate from the initial compressed and
out-of-equilibrium phase of the collision.
For the isospin effects the neutron-proton $differential$ flows
$
V^{(n-p)}_{1,2} (y,p_t) \equiv V^n_{1,2}(y,p_t) - V^p_{1,2}(y,p_t)
$ 
have been suggested as very useful probes of the isovector part of 
the $EoS$ since they appear rather insensitive to the isoscalar potential
and to the in medium nuclear cross sections, \cite{BaoPRL82,BaoPRL85}.

In heavy-ion collisions around $1AGeV$ with
radioactive beams,
 due to the large 
counterstreaming nuclear currents, one may exploit the
different Lorentz nature of a scalar and a vector field, see the different
$\gamma$-boosting in the local force, Eq.(\ref{force}).
In Fig.\ref{flows}
transverse and elliptic differential flows are shown
for 
the $^{132}Sn+^{124}Sn$
reaction at $1.5~AGeV$ ($b=6fm$), that likely could be studied with 
the new planned radioactive beam facilities at intermediate energies,
 \cite{GrecoPLB562}. 
The effect of the different structure of the 
isovector channel is clear. Particularly evident is the splitting in 
the high $p_t$
region of the elliptic flow.
 In the $(\rho+\delta)$ dynamics the high-$p_t$ neutrons show a much larger 
$squeeze-out$.
This is fully consistent with an early emission (more spectator shadowing)
due to the larger repulsive $\rho$-field.
The $V_2$ observable, which is a good {\it chronometer} of the reaction
dynamics, appears to be particularly sensitive to the Lorentz structure
of the effective interaction. We expect similar effects, even enhanced, from 
the measurements of 
differential flows for light isobars, like $^3H~vs.~^3He$.

\section{Isospin effects on sub-threshold kaon production at intermediate 
energies} 

Particle production represent a  
useful tool to constrain the poorly known high density behaviour of the  
nuclear equation of state ($EoS$) \cite{StockPR135,AichkoPRL55}. 
In particular pion and (subthreshold) kaon productions have been extensively 
investigated  leading to the conclusion of 
a soft 
behaviour of the $EoS$ 
at high densities, \cite{FuchsPPNP56,HartPRL96}. 
Kaons ($K^0,K^+$)
appear as particularly sensitive probes since they are produced in the high 
density phase almost without subsequent reabsorption effects 
\cite{FuchsPPNP56}.
At variance, 
antikaons ($\bar K^0,\bar K^-$) are strongly coupled to the hadronic medium 
through strangeness
exchange reactions \cite{FuchsPPNP56,CassingNPA727}.
Here we show that the isospin dependence of the $K^{0,+}$
production can be also used to probe the isovector part of
the $EoS$: we propose the $K^0/K^+$ yield ratio as a good observable to
constrain the high density behavior of the symmetry energy,
 $E_{sym}$, \cite{bao,baranPR}. 

Using our $RMF$ transport model  we analyze 
pion and kaon production in central $^{197}Au+^{197}Au$ collisions in 
the $0.8-1.8~AGeV$
 beam 
energy range, with different effective field choices for 
$E_{sym}$. We will compare results of three Lagrangians with constant 
nucleon-meson 
couplings ($NL...$ type, see before) and one with density
dependent couplings ($DDF$, see \cite{gait04}), recently suggested 
for better nucleonic properties of neutron stars \cite{Klahn06}.
In the $DDF$ model
the $f_{\rho}$ is exponentially decreasing with density, resulting in a 
rather "soft" 
symmetry term at high density. 
In order to isolate the sensitivity to the isovector components we use
models showing the same "soft" $EoS$ for symmetric matter.

Pions are produced via the decay of the $\Delta(1232)$ resonance and 
can contribute 
to the kaon yield through 
collisions with baryons: 
$\pi B \longrightarrow YK$. All these processes are treated within
 a relativistic transport model including an hadron mean field propagation.
The latter point, which goes beyond the ``collision cascade'' picture, is
essential for particle production yields since it directly affects the
energy balance of the inelastic channels.

\begin{figure}[t] 
\begin{center}
\includegraphics[scale=0.32]{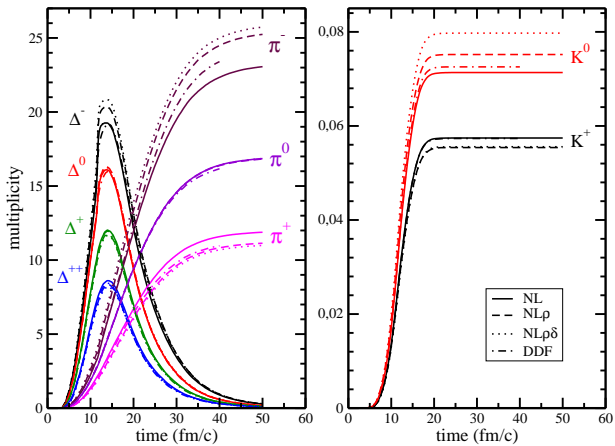} 
\caption{Time evolution of the $\Delta^{\pm,0,++}$ resonances and  
pions $\pi^{\pm,0}$ (left), and kaons $K^{0,+}$ (right) for a central 
($b=0$ fm impact parameter).  
Au+Au collision at 1 AGeV incident energy. Transport calculation using the  
$NL, NL\rho, NL\rho\delta$ and $DDF$ models for the iso-vector part of the  
nuclear $EoS$ are shown.  
}
\label{kaon1} 
\end{center}
\end{figure}

Fig. \ref{kaon1} reports  the temporal evolution of $\Delta^{\pm,0,++}$  
resonances and pions ($\pi^{\pm,0}$) and kaons ($K^{+,0}$)  
for central Au+Au collisions at $1~AGeV$.
It is clear that, while the pion yield freezes out at times of the order of 
$50 fm/c$, i.e. at the final stage of the reaction (and at low densities),
kaon production occur within the very early stage of the reaction,
and the yield saturates at around $20 fm/c$. Kaons are then suitable 
to probe the 
high density phase of nuclear matter. 
From Fig. \ref{kaon1} we see that the pion results are  
weakly dependent on the  
isospin part of the nuclear mean field.
However, a slight increase (decrease) in the $\pi^{-}$ ($\pi^{+}$) 
multiplicity is observed when going from the $NL$ (or $DDF$) to the 
$NL\rho$ and then to
the $NL\rho\delta$ model, i.e. increasing the vector contribution $f_\rho$
in the isovector channel. This trend is 
more pronounced for kaons, see the
right panel, due to the high density selection of the source and the
proximity to the production threshold. 
 The results for the $DDF$ model,
 density dependent couplings with a large $f_{\rho}$ decrease at high density,
 are fully consistent. They are always closer to the $NL$ case (without
 isovector interactions) but the difference, still seen for $\pi^{+,-}$,
 is completely disappearing for $K^{0,+}$, selectively produced at high 
 densities. 

When isovector fields are included the symmetry potential energy in 
neutron-rich matter is repulsive for neutrons and attractive for protons.
In a $HIC$ this leads to a fast, pre-equilibrium, emission of neutrons.
 Such a $mean~field$ mechanism, often referred to as isospin fractionation
\cite{bao,baranPR}, is responsible for a reduction of the neutron
to proton ratio during the high density phase, with direct consequences
on particle production in inelastic $NN$ collisions.

$Threshold$ effects represent a more subtle point. The energy 
conservation in
a hadron collision in general has to be formulated in terms of the canonical
momenta, i.e. for a reaction $1+2 \rightarrow 3+4$ as
$
s_{in} = (k_1^\mu + k_2^\mu)^2 = (k_3^\mu + k_4^\mu)^2 = s_{out}.
$
Since hadrons are propagating with effective (kinetic) momenta and masses,
 an equivalent relation should be formulated starting from the effective
in-medium quantities $k^{*\mu}=k^\mu-\Sigma^\mu$ and $m^*=m+\Sigma_s$, where
$\Sigma_s$ and $\Sigma^\mu$ are the scalar and vector self-energies,
Eqs.(\ref{selfen}).
In reactions where nucleon resonances, especially the different isospin
states of the $\Delta$ resonance, and hyperons enter, also their self
energies are relevant for energy conservation. 
We specify them in the usual
way according to the light quark content and with appropriate Clebsch-Gordon
coefficients \cite{FeriniNPA762}. 
The self-energy contribution to the energy conservation in inelastic 
channels will
influence the particle production in two different ways. On one hand it will 
directly determine the thresholds and thus the multiplicities of a certain type
of particles, in particular of the sub-threshold ones, as 
here for the kaons. Secondly it will modify the phase space 
available in the final channel.

\begin{figure}[t] 
\begin{center}
\includegraphics[scale=0.33]{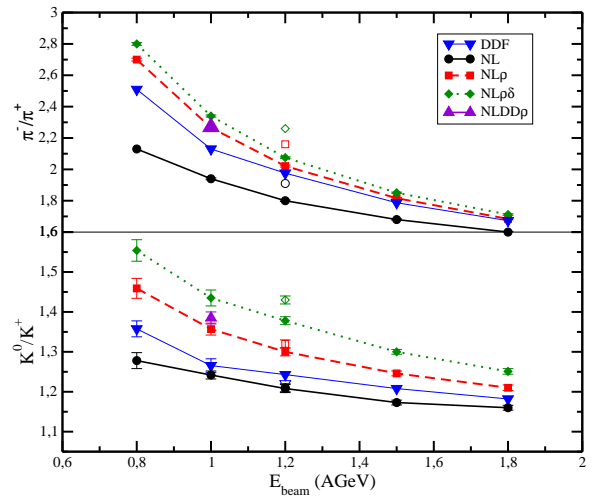}
\caption{
$\pi^-/\pi^+$ (upper) and  $K^{+}/K^{0}$ (lower) ratios
as a function of the incident energy for the same reaction and models
as in Fig. \ref{kaon1}.
In addition we present, for $E_{beam}=1~AGeV$, $NL\rho$ results with a density 
dependent
$\rho$-coupling (triangles), see text. The $open$ symbols
at $1.2~AGeV$ show the corresponding results for a $^{132}Sn+^{124}Sn$
collision, more neutron rich.
}
\label{kaon2} 
\end{center}
\end{figure} 
 
In fact in neutron-rich systems {\it mean field} 
and {\it threshold} effects
are acting in opposite directions on particle production  and might 
compensate each other.
 As an example, $nn$
collisions excite $\Delta^{-,0}$ resonances which decay mainly to $\pi^-$.
 In a
neutron-rich matter the mean field effect pushes out neutrons making the 
matter more symmetric and thus decreasing the $\pi^-$ yield. The threshold 
effect on the other hand is increasing the rate of $\pi^-$'s due to the
enhanced production of the $\Delta^-$ resonances: 
now the $nn \rightarrow p\Delta^-$ process is favored
(with respect to $pp \rightarrow n\Delta^{++}$) 
 since more effectively a neutron is converted into a proton.
Such interplay between the two mechanisms cannot be 
fully included in a non-relativistic dynamics,
in particular in calculations where the baryon symmetry potential is
treated classically \cite{BaoPRC71,QLiPRC72}.

In the $0.8-1.8~AGeV$ range the sensitivity
is larger for the $K^{0}/K^{+}$ compared to the $\pi^-/\pi^+$
ratio, as we can see from Fig.\ref{kaon2}. 
This is due to the
subthreshold production and to the fact that
 isospin effects enter twice in the two-steps production of kaons, 
\cite{twosteps}. 
Between the two extreme $DDF$ and
$NL\rho\delta$ isovector interaction models, the 
variations in the ratios are of the order of $14-16 \%$ for kaons, while 
they reduce to about $8-10 \%$ for pions.
Interestingly the Iso-$EoS$ effect for pions is increasing at lower energies,
when approaching the production threshold.
We have to note that in a previous study of kaon production in excited nuclear
matter the dependence of the $K^{0}/K^{+}$ yield ratio on the effective
isovector interaction appears much larger (see Fig.8 of 
ref.\cite{FeriniNPA762}).
The point is that in the non-equilibrium case of a heavy ion collision
the asymmetry of the source where kaons are produced is in fact reduced
by the $n \rightarrow p$ ``transformation'', due to the favored 
$nn \rightarrow p\Delta^-$ processes. This effect is almost absent at 
equilibrium due to the inverse transitions, see Fig.3 of 
ref.\cite{FeriniNPA762}. Moreover in infinite nuclear matter even the fast
neutron emission is not present. 
This result clearly shows that chemical equilibrium models can lead to
uncorrect results when used for transient states of an $open$ system.
In the same Fig. \ref{kaon2} we also report results at $1.2~AGeV$ for
the $^{132}Sn+^{124}Sn$ reaction, induced by a radioactive beam, with
an overall larger asymmetry (open symbols). The isospin effects are 
clearly enhanced.

\section{Testing Deconfinement at High Isospin Density}
The hadronic matter is expected to undergo a phase transition 
into a deconfined phase of quarks and gluons at large densities 
and/or high temperatures. On very general grounds,
the transition's critical densities are expected to depend
on the isospin of the system, but no experimental tests of this 
dependence have been performed so far.
Moreover, up to now, data on the phase transition have been 
extracted from
ultrarelativistic collisions, when large temperatures but low baryon densities
are reached. 
\begin{figure}
\begin{center}
\includegraphics[scale=0.36]{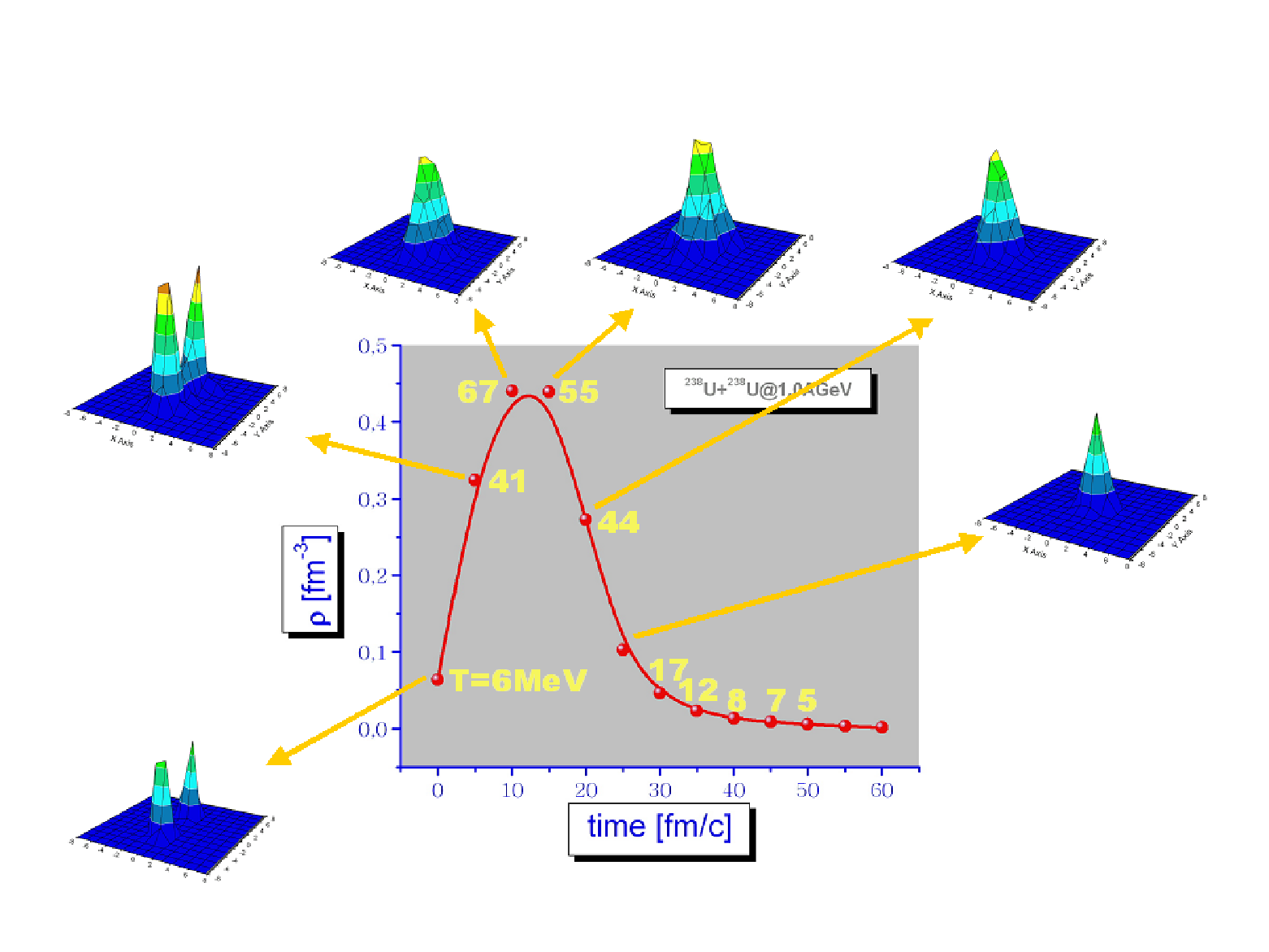}
\vskip -1.3cm
\includegraphics[angle=-90,scale=0.30,trim=0 0 -20 0]{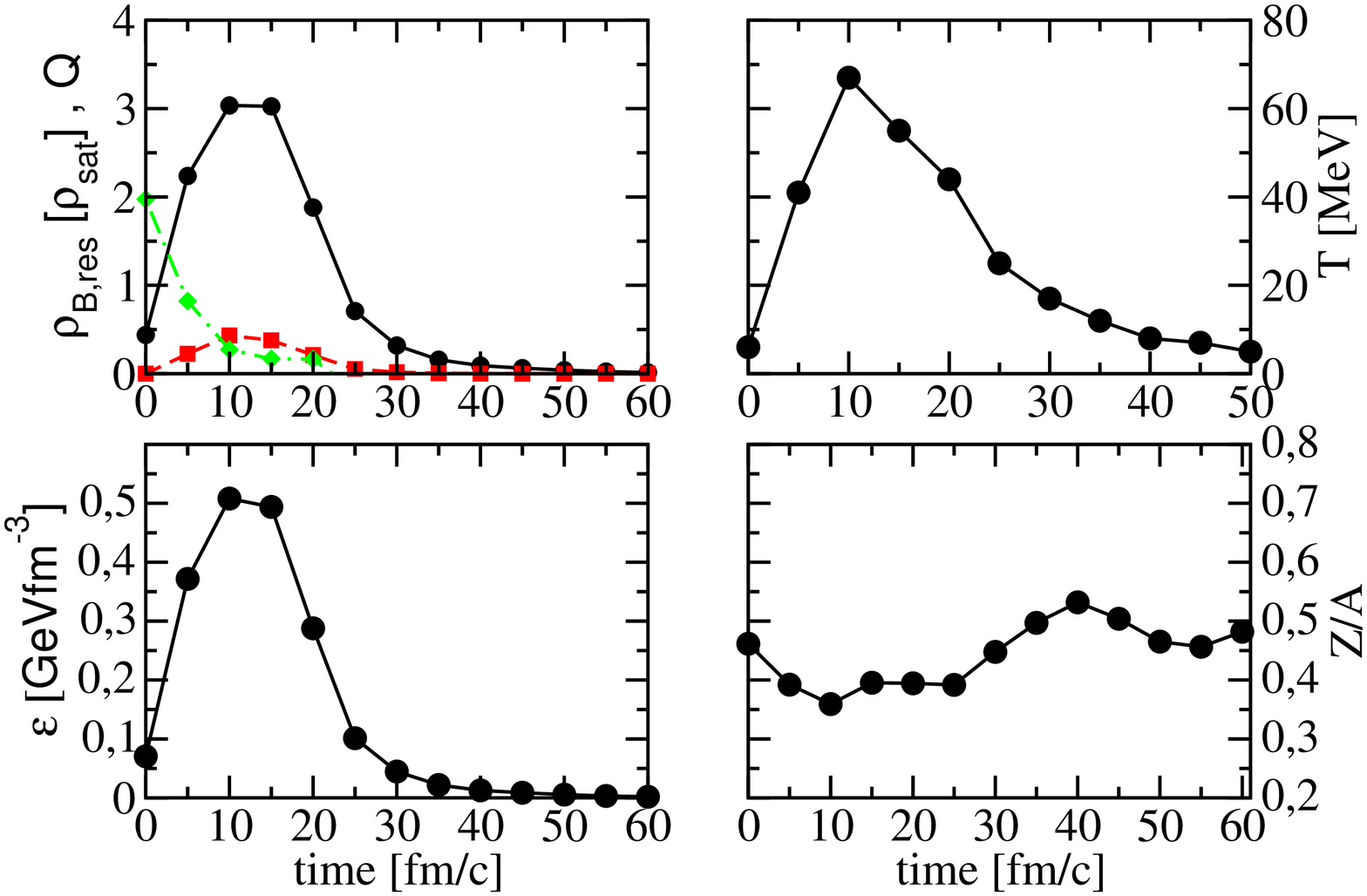}
\caption{\label{figUU}
Uranium-Uranium $1~AGeV$ semicentral. Correlation between 
density, 
temperature, momentum
thermalization inside a cubic cell 2.5 $fm$ wide, located in the center
of mass of the system.
 Lower panel: density, temperature, energy density, 
momentum, proton fraction. Curves in the upper-left: {\it black dots} -
baryon density in $\rho_0$ units; {\it grey dots} - quadrupole moment in 
momentum space; {\it squares} - resonance density.    
}
\vskip -1.7cm
\end{center}
\end{figure}
In order to check the possibility of observing some precursor signals
of some new physics even in collisions of stable nuclei at
intermediate energies we have performed some event simulations for the
collision of very heavy, neutron-rich, elements. We have chosen the
reaction $^{238}U+^{238}U$ (average proton fraction $Z/A=0.39$) at
$1~AGeV$ and semicentral impact parameter $b=7~fm$ just to increase
the neutron excess in the interacting region. 
 To
evaluate the degree of local equilibration and the corresponding
temperature we have followed the momentum distribution in a space
cell located in the c.m. of the system; in the same cell we report the
maximum mass density evolution. Results are shown in  Fig.~\ref{figUU}.  
We see that after about $10~fm/c$ a nice local
equilibration is achieved.  We have a unique Fermi distribution and
from a simple fit we can evaluate the local temperature.
  At this beam energy the maximum density 
 coincides with the thermalization, then the system is quickly cooling 
 while expanding.
 In Fig.\ref{figUU}, lower panel, we report the time evolution of all physics
 parameters inside the c.m. cell in the interaction region..
We note that a rather exotic nuclear matter is formed in a transient
time of the order of $10~fm/c$, with baryon density around $3-4\rho_0$,
temperature $50-60~MeV$, energy density $500~MeV~fm^{-3}$ and proton
fraction between $0.35$ and $0.40$, well inside the estimated mixed 
phase region, see the following..

A study of the isospin dependence of the transition densities has been
performed up to now, to our knowledge, only by Mueller
\cite{MuellerNPA618}.  The conclusion  is that,
moving from symmetric nuclei to nuclei having $Z/A\sim 0.3$, the
critical density is reduced by roughly 10$\%$.  Here we
explore in a more systematic way the model parameters and we estimate
the possibility of forming a mixed-phase of quarks and hadrons in
experiments at energies of the order of a few $GeV$ per nucleon.

Concerning the hadronic phase, we use the relativistic
non-linear Walecka-type model of Glendenning-Moszkowski ($GM...$) 
\cite{GlendenningPRL18}, where the isovector part is treated 
just with $\rho$-meson couplings, and
the iso-stiffer $NL\rho\delta$ interaction \cite{deconf06}. 

For the quark phase we consider the $MIT$ bag model
\cite{MitbagPRD9} with various bag pressure constants.  In particular 
we are interested in those parameter sets
which would allow the existence of quark stars
\cite{HaenselAA160,DragoPLB511}, i.e. parameters sets for
which the so-called Witten-Bodmer hypothesis is satisfied
\cite{WittenPRD30,BodmerPRD4}. 
One of the
aim of our work it to show that if quark stars are indeed possible,
it is then very likely to find signals of the formation of a mixed
quark-hadron phase in intermediate-energy heavy-ion experiments
\cite{deconf06}.

The
scenario we would like to explore corresponds to the situation
realized in experiments at moderate energy, in which the temperature
of the system is at maximum of the order of a few ten $MeV$.  In this
situation,
only a tiny amount of strangeness can be produced and therefore 
we can only study the deconfinement transition from hadron
matter into up and down quark matter. Since there no time for weak decays
the environment is rather different from the neutron star case.

\subsubsection*{Mixed phase}
The structure of the mixed phase is obtained by
imposing the Gibbs conditions \cite{Landaustat,GlendenningPRD46} for
chemical potentials and pressure and by requiring
the conservation of the total baryon and isospin densities
\begin{eqnarray}\label{gibbs}
&&\mu_B^{(H)} = \mu_B^{(Q)}\, ,~~  
\mu_3^{(H)} = \mu_3^{(Q)} \, ,\nonumber \\  
&&P^{(H)}(T,\mu_{B,3}^{(H)}) = P^{(Q)} (T,\mu_{B,3}^{(Q)})\, ,\nonumber \\
&&\rho_B=(1-\chi)\rho_B^H+\chi\rho_B^Q \, ,\nonumber \\
&&\rho_3=(1-\chi)\rho_3^H+\chi\rho_3^Q\, , 
\end{eqnarray}
where $\chi$ is the fraction of quark matter in the mixed phase.
In this way we get the $binodal$ surface which gives the phase coexistence 
region
in the $(T,\rho_B,\rho_3)$ space
\cite{GlendenningPRD46,MuellerNPA618}. For a fixed value of the
conserved charge $\rho_3$, 
 related to the proton fraction $Z/A \equiv (1+\rho_3/\rho_B)/2$, 
 we will study the boundaries of the mixed phase
region in the $(T,\rho_B)$ plane. We are particularly interested in
the lower baryon density border, i.e. the critical/transition density
$\rho_{cr}$, in order to check the possibility of reaching such
$(T,\rho_{cr},\rho_3)$ conditions in a transient state during an $HIC$
at relativistic energies.
In the hadronic phase the charge chemical potential is given by
$
\mu_3 = 2 E_{sym}(\rho_B) \frac{\rho_3}{\rho_B}\, .
$ 
Thus, we expect critical densities
rather sensitive to the isovector channel in the hadronic $EoS$.

\begin{figure}
\begin{center}
\includegraphics[angle=+90,scale=0.37]{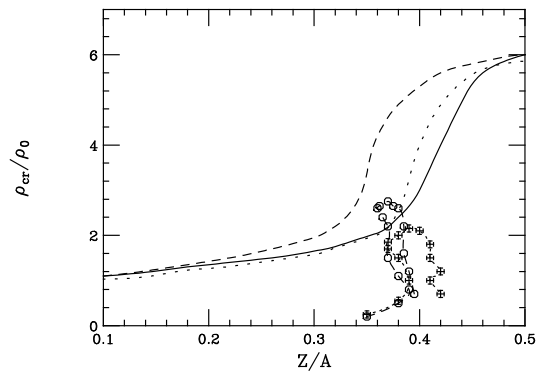}
\caption{\label{rhodelta}
Variation of the transition density with proton fraction for various
hadronic $EoS$ parameterizations. Dotted line: $GM3$ parametrization;
 dashed line: $NL\rho$ parametrization; solid line: $NL\rho\delta$ 
parametrization. For the quark $EoS$, the $MIT$ bag model with
$B^{1/4}$=150 $MeV$.
The points represent the path followed
in the interaction zone during a semi-central $^{132}$Sn+$^{132}$Sn
collision at $1~AGeV$ (circles) and at $300~AMeV$ (crosses), see text. 
}
\vskip -1.5cm
\end{center}
\end{figure}

We compare the predictions on the transition to a
deconfined phase of the two effective Lagrangians $GM3$ and
$NL\rho\delta$.  The isoscalar part is very similar while the isovector 
$EoS$ is different, because in $GM3$ we only have the coupling to the vector
$\rho$-field.
In Fig.~\ref{rhodelta}  we show the crossing
density $\rho_{cr}$ separating nuclear matter from the quark-nucleon
mixed phase, as a function of the proton fraction $Z/A$.  
We can see the effect of the
$\delta$-coupling towards an $earlier$ crossing due to the larger
symmetry repulsion at high baryon densities. 
The $\delta$-exchange potential provides an extra isospin repulsion
of the hadron $EoS$, and its effect shows up
in a further reduction of the
critical density.

In the same figure we report the paths in the $(\rho,Z/A)$
plane followed in the c.m. region during the collision of the n-rich
 $^{132}$Sn+$^{132}$Sn system, at different energies. We see that already at
$300~AMeV$ we are reaching the border of the mixed phase, and we are
well inside it at $1~AGeV$. 

\subsubsection*{Deconfinement Precursors}
Statistical
fluctuations could help in  reducing the density at which drops of quark
matter form. The reason is that a small bubble can
be energetically favored if it contains quarks whose Z/A ratio is
{\it smaller} than the average value of the surrounding region. This
is again due to the strong Z/A dependence of the free energy, which
favors clusters having a small electric charge. 
These configurations can easily transform into a
bubble of quarks having the same flavor content of the original
hadrons, even if the density of the system is not large enough to
allow deconfinement in the absence of statistical fluctuations.  
Moreover, since 
fluctuations favor the formation of bubbles having a smaller Z/A,
neutron emission from the central collision area should be suppressed,
what could give origin to specific signatures of the mechanism
described in this paper. This corresponds to a {\it neutron trapping}
effect, supported also by a symmetry energy difference in the
two phases.
In fact while in the hadron phase we have a large neutron
potential repulsion (in particular in the $NL\rho\delta$ case), in the
quark phase we only have the much smaller kinetic contribution.
 If in a
pure hadronic phase neutrons are quickly emitted or ``transformed'' in
protons by inelastic collisions, when the mixed phase
starts forming, neutrons are kept in the interacting system up to the
subsequent hadronization in the expansion stage \cite{deconf06}.
Observables related to such neutron ``trapping'' could be an
inversion in the trend of the formation of neutron rich fragments
and/or of the $\pi^-/\pi^+$, $K^0/K^+$ yield ratios for reaction
products coming from high density regions, i.e. with large transverse
momenta.  In general we would expect a modification of the rapidity
distribution of the emitted ``isospin'', with an enhancement at
mid-rapidity joint to large event by event fluctuations.
 A more detailed analysis is clearly needed.

\section{Perspectives}
We have shown in few examples the richness of the physics we can describe
using mean field transport equations, inspired by the pioneering works
of David M. Brink.
We have seen that collisions of n-rich heavy ions from low to 
intermediate energies
can bring new information on the isovector part of the in-medium interaction
in different regions of high baryon densities, qualitatively different 
from equilibrium
$EoS$ properties. We have presented quantitative results for 
charge equilibration in fusion/deep inelastic reactions, differential
collective flows and yields of charged pion and kaon ratios.
Important non-equilibrium effects for particle production are stressed.
Finally our study supports the possibility of observing
precursor signals of the phase transition to a mixed hadron-quark matter
at high baryon density in the collision, central or semi-central, of
neutron-rich heavy ions in the energy range of a few $GeV$ per
nucleon.  As signatures we 
suggest to look at observables particularly sensitive to the
expected different isospin content of the two phases, which leads to a
neutron trapping in the quark clusters.
The isospin structure of hadrons produced at high transverse momentum
should be a good indicator of the effect.
 
Many new ideas for fundamental experiments with radioactive beams
are emerging. My picture of David Brink like a tree always starting new
braches is more and more confirmed......

\vskip 1.0cm

{\it Acknowledgements}

\noindent
I would like to warmly mention the great experience of collaborating with
exceptional people on the topics and projects shortly discussed here. I will 
try to list some of them: V.Baran, M.Colonna, A.Drago, T.Gaitanos, V.Greco, 
A.Lavagno, B.Liu, M.Pfabe, H.H.Wolter, S.Yildirim and the essential young 
contributors  L.Bonanno, G.Ferini, R.Lionti, N.Pellegriti, V.Prassa, J.Rizzo 
and E.Santini.

The interaction with experimental groups has been essential, in particular 
I like to thank the DIPOLE Collab. (D.Pierroutsakou et al.), the 
CHIMERA Collab. (A.Pagano et al.), and the FOPI Collab. (W.Reisdorf et al.)
for the intense discussions and the access to the data.




\begin{thebibliography}{00}

\bibitem{bao}
B.A. Li, W.U. Schroeder (Eds.), Isospin Physics in Heavy-Ion Collisions 
at Intermediate Energies, Nova Science, New York, 2001.

\bibitem{baranPR} V.Baran, M.Colonna, V.Greco, M.Di Toro, 
{\em Phys. Rep.} {\bf 410} (2005) 335.

\bibitem{ChomazNPA563}
P. Chomaz, M. Di Toro, A.Smerzi,
{\em Nucl. Phys.} {\bf A563} (1993) 509.

\bibitem{BortignonNPA583}
P. F. Bortignon et al.,
{\em Nucl. Phys.} {\bf A583} (1995) 101c.

\bibitem{FlibPRL77}
S. Flibotte et al.,
{\em Phys. Rev. Lett.} {\bf 77} (1996) 1448.

\bibitem{CinNC111}
M. Cinausero et al.,
{\em Nuovo Cimento} {\bf 111} (1998) 613.

\bibitem{PierrouEPJA16}
D. Pierroutsakou et al.,
{\em Eur. Phys. Jour.} {\bf A16} (2003) 423, 
 {\em Nucl. Phys.} {\bf A687} (2003) 245c.

\bibitem{AmoPRC29}
F.Amorini et al.,
{\em Phys. Rev.} {\bf C69} (2004) 014608.

\bibitem{PierrouPRC71}
D. Pierroutsakou et al.,
{\em Phys. Rev.} {\bf C71} (2005) 054605.

\bibitem{BrinkNPA372}
D. M. Brink and M. Di Toro,,
{\em Nucl. Phys.} {\bf A372} (1981) 151.

\bibitem{BaranNPA600}
V. Baran et al., 
{\em Nucl. Phys.} \textbf{A600} (1996) 111.

\bibitem{CabNPA637}
M.Cabibbo et al.,
{\em Nucl. Phys.} \textbf{A637} (1998) 374. 

\bibitem{BaranNPA679}
V. Baran et al., 
{\em Nucl. Phys.} \textbf{A679} (2001) 373.

\bibitem{SimenPRL86}
C. Simenel et al.,
{\em  Phys. Rev. Lett.} \textbf{86} (2000) 2971.

\bibitem{BaranPRL87}
V. Baran, D. M. Brink, M. Colonna, M. Di Toro, 
{\em Phys. Rev. Lett.} {\bf 87} (2001) 182501

\bibitem{PierrouLNS}
D.Pierroutsakou et al.,
LNS  exp. proposal  2006, PAC approved.

\bibitem{luca04}
L. Bonanno,  
{\em Effetti di radiazione diretta dipolare sulla sintesi degli elementi
 superpesanti}, Master Thesis, Catania Univ. 2004.

\bibitem{spiral2}
A letter of intent for the new SPIRAL2 facility at GANIL is in preparation.

\bibitem{qhd}
B. D. Serot, J. D. Walecka, 
Advances in  Nuclear Physics, {\bf 16}, 1,
eds. J. W. Negele, E. Vogt, (Plenum, N.Y., 1986).

\bibitem{KoPRL59}
C. M. Ko, Q. Li, R. Wang,
{\em Phys. Rev. Lett.} {\bf 59} (1987) 1084.

\bibitem{GiessenRPP56}
B. Bl\"attel, V. Koch, U. Mosel, 
{\em Rep. Prog. Phys.} {\bf 56} (1993) 1.

\bibitem{kubis}
S. Kubis, M. Kutschera, 
{\em Phys. Lett.} {\bf B399} (1997) 191.

\bibitem{liu}
B. Liu, V. Greco, V. Baran, M. Colonna, M. Di Toro, 
{\em Phys. Rev.} {\bf C65} (2002) 045201.

\bibitem{gait04}
T. Gaitanos, M. Di Toro, S. Typel, V. Baran, C. Fuchs, V. Greco, H.H. Wolter, 
{\em Nucl. Phys.} {\bf A732} (2004) 24.

\bibitem{dbhf}
F. de Jong, H. Lenske, 
{\em Phys. Rev.} {\bf C57} (1998) 3099;\\
E.N.E. van Dalen, C. Fuchs, A. Faessler, 
{\em Nucl. Phys.} {\bf A744} (2004) 227.

\bibitem{diracmass}
E.N.E. van Dalen, C. Fuchs, A. Faessler, 
{\em Phys. Rev. Lett.} {\bf 95} (2005) 022302.

\bibitem{msuria}
M. Di Toro, M. Colonna, J. Rizzo,
AIP Conf. Proc., Vol.{\bf 791} (2005) 70-83.

\bibitem{RizzoPRC72}
J. Rizzo, M. Colonna, M. Di Toro,
{\em Phys. Rev.} {\bf C72} (2005) 064609. 

\bibitem{dani}
P. Danielewicz, Roy A. Lacey, et al., 
{\em Phys. Rev. Lett.} {\bf 81}, (1998) 2438; \\
C. Pinkenburg {\it et al.}, 
{\em Phys. Rev. Lett.} {\bf 83}, (1999) 1295.


\bibitem{gait01}
T.~Gaitanos, C.~Fuchs, H.H.~Wolter, A. Faessler, 
Eur. Phys. J. A 12 (2001) 421;\\
T. Gaitanos, C. Fuchs, H.H. Wolter, 
{\em Nucl. Phys.} {\bf A741} (2004) 287.

\bibitem{FuchsNPA589}
C. Fuchs. H.H. Wolter,
{\em Nucl. Phys.} {\bf A589} (1995) 732.

\bibitem{huber}
H. Huber, J. Aichelin, 
{\em Nucl. Phys.} {\bf A573} (1994) 587.

\bibitem{FeriniNPA762}
G. Ferini, M. Colonna, T. Gaitanos, M. Di Toro,
{\em Nucl. Phys.} {\bf A762} (2005) 147.

\bibitem{GrecoPLB562}
V. Greco, V. Baran, M. Colonna, M. Di Toro, T. Gaitanos, 
H.H. Wolter, 
{\em Phys. Lett.} {\bf B562} (2003) 215.

\bibitem{OlliPRD46} 
J.Y. Ollitrault, 
{\em Phys. Rev.} {\bf D46} (1992) 229.

\bibitem{DanielNPA673} 
P. Danielewicz, 
{\em Nucl. Phys.} {\bf A673} (2000) 375.

\bibitem{BaoPRL82} 
B.A. Li and A.T. Sustich, 
{\em Phys. Rev. Lett.} {\bf 82} (1999) 5004.

\bibitem{BaoPRL85}
B.A. Li, 
{\em Phys. Rev. Lett.} {\bf 85} (2000) 4221.

\bibitem{StockPR135}
R. Stock, 
{\em Phys. Rep.} {\bf 135} (1986) 259.

\bibitem{AichkoPRL55} 
J. Aichelin, C.M. Ko,  
{\em Phys. Rev. Lett.} {\bf 55} (1985) 2661. 

\bibitem{FuchsPPNP56}
C. Fuchs, {\em Prog.Part.Nucl.Phys.} {\bf 56} 1-103 (2006). 

\bibitem{HartPRL96}
C.Hartnack, H.Oeschler, J.Aichelin,
{\em Phys. Rev. Lett.} {\bf 96} (2006) 012302. 

\bibitem{CassingNPA727}
W. Cassing, L.Tolos, E.L. Bratkovskaya, A. Ramos,
{\em Nucl. Phys.} {\bf A727} 59 (2003).

\bibitem{WeberPRC67}
H. Weber, E.L. Bratkovskaya, W. Cassing, H. St\"ocker,
 {\em Phys. Rev.} {\bf C67} 014904 (2003).

\bibitem{Klahn06}
T.Kl\"ahn et al. {\em Constraints on the high-density nuclear equation of state
...}, {\em arXiv:nucl-th/0602038}. 

\bibitem{BaoPRC71} 
B.A. Li, G.C. Yong, W. Zuo,  
{\em Phys. Rev.} {\bf C71} 014608 (2005). 
  
\bibitem{QLiPRC72}
Q. Li et al., 
{\em Phys. Rev.} {\bf C72} 034613 (2005).

\bibitem{twosteps}
In the energy 
range explored here, the main contribution to the kaon yield comes from the
pionic channels, in particular from $\pi N$ collisions, and from the
 $N \Delta$ channel, which together account 
for nearly $80 \%$ of the total yield, see \cite{FeriniNPA762}. 

\bibitem{MuellerNPA618} H.Mueller, 
{\em Nucl. Phys.} {\bf A618} (1997) 349.

\bibitem{GlendenningPRL18} N.K.Glendenning, S.A.Moszkowski, 
{\em Phys. Rev. Lett.} {\bf 67} (1991) 2414.

\bibitem{deconf06}
M. Di Toro, A. Drago, T. Gaitanos, V. Greco, A. Lavagno,
{\em Testing Deconfinement...}, {\em nucl-th/0602052}, 
{\em Nucl. Phys.} {\bf A} (2006) in press. 

\bibitem{MitbagPRD9} A.Chodos et al., 
{\em Phys. Rev.} {\bf D9} (1974) 3471.

\bibitem{HaenselAA160} P.Haensel, J.L.Zdunik, R.Schaeffer, 
{\em Astron. Astrophys.} {\bf 160} (1986) 121.

\bibitem{DragoPLB511} A.Drago, A.Lavagno, 
{\em Phys. Lett.} {\bf B511} (2001) 229.

\bibitem{WittenPRD30} E.Witten, 
{\em Phys. Rev.} {\bf D30} (1984) 272.

\bibitem{BodmerPRD4} A.R.Bodmer, 
{\em Phys. Rev.} {\bf D4} (1971) 1601.              
 
\bibitem{AlfordPRD67} M.Alford, S.Reddy, 
{\em Phys. Rev.} {\bf D67} (2003) 074024.

\bibitem{DragoPRD69} A.Drago, A.Lavagno, G.Pagliara, 
Phys.Rev. {\bf D69} (2004) 057505.

\bibitem{Landaustat} L.D.Landau, L.Lifshitz, {\em Statistical Physics},
 Pergamon Press, Oxford 1969.

\bibitem{GlendenningPRD46} N.K.Glendenning, 
{\em Phys. Rev.} {\bf D46} (1992) 1274.


\end{thebibliography}
\end{document}